\documentclass[12pt,a4paper,fleqn]{article}
\usepackage{amsmath}
\usepackage{graphicx}
\usepackage{cite}
\usepackage{multirow}

\usepackage[american]{babel}

\usepackage{dcolumn} 
\newcolumntype{d}{D{.}{.}{2}}
\newcolumntype{e}{D{.}{.}{3}}
\newcolumntype{f}{D{.}{.}{4}}

\usepackage{endfloat} 
\AtBeginDelayedFloats{\linespread{1.66}}

\begin{document}

\begin{center}

{\LARGE\bf
Construction of CASCI-type wave functions 
for very large active spaces
}

{\large 
Katharina Boguslawski,
Konrad H.\ Marti,
and Markus Reiher\footnote{Author to whom correspondence should be sent; email: markus.reiher@phys.chem.ethz.ch, FAX: ++41-44-63-31594, TEL: ++41-44-63-34308}
}\\[2ex]

ETH Zurich, Laboratorium f{\"u}r Physikalische Chemie, 
Wolfgang-Pauli-Str.\ 10,\\
CH-8093 Zurich, Switzerland \\[2ex]

\end{center}

\begin{center}
{\large\bf Abstract}\\[0.2ex]
\end{center}

{\small
We present a procedure to construct a configuration-interaction 
expansion containing arbitrary excitations from an underlying full-configuration-interaction-type
wave function defined for a very large active space. 
Our procedure is based on the density-matrix renormalization group (DMRG) algorithm
that provides the necessary information in terms of the eigenstates of the
reduced density matrices to calculate the coefficient of any basis state in the
many-particle Hilbert space.
Since the dimension of the Hilbert space scales binomially with the size of
the active space, a sophisticated Monte Carlo sampling routine is employed. 
This sampling algorithm can also construct such configuration-interaction-type wave functions from
any other type of tensor network states. 
The configuration-interaction information obtained serves several purposes. It yields a qualitatively correct
description of the molecule's electronic structure, it allows us to analyze DMRG wave functions converged
for the same molecular system but with different parameter sets (e.g., different numbers of active-system (block) states),
and it can be considered a balanced reference for the application of a subsequent standard multi-reference
configuration-interaction method. 

\vfil

\begin{tabbing}
Date:   \quad \= May 12, 2011  \\
Status:       \> revised version accepted by \textit{J. Chem. Phys.}\\
\end{tabbing}

\newpage

\section{Introduction}
In quantum chemistry, electronic wave functions are usually represented 
as configuration-interaction (CI) expansions, i.e., wave functions expanded in a set of Slater
determinants or configuration state functions \cite{jorgensen}. 
The dimension of this many-particle basis, however, grows binomially with the number of orbitals
and electrons present in the molecule under study which restricts the application of CI-type methods.
In 1990, Olsen, J{\o}rgensen and coworkers managed to perform full CI (FCI) calculations with more than one billion
Slater determinants \cite{Olsen1990}. 

It was found that CI vectors are actually sparse \cite{knowles} if contributions below a predefined threshold are neglected.
Different approaches have been proposed which take advantage of the sparsity of CI vectors to either
overcome the limitations in FCI calculations \cite{knowles_fci,harrison,luzanov}
or to systematically determine only the important configurations in the wave function expansion
\cite{Mitrushenkov,feller,greer}. In particular, the latter procedures allow a dynamic selection of
important configurations without any predefined and limiting excitation scheme applied to a reference
configuration \cite{greer98}. Such methods are, in principle, applicable to any kind of electronic structure.
For specific types, like {\it closed-shell} electronic structures, special techniques have been developed as powerful
and efficient tools like the standard single-reference coupled-cluster (CC) method \cite{jorgensen}, which generates
a predefined CI expansion based on a specific orbital-excitation hierarchy.
Still, generally applicable CI methods are sought for difficult cases like for open-shell molecules with a multi-reference character
(for which open-shell transition metal clusters are an example \cite{markus_chimia_2009,marti2010pccp,markus_fd2}) or for excited states.

Such a promising method is the density matrix renormalization group (DMRG) algorithm 
introduced by White into solid state physics\cite{white,PhysRevLett.68.3487}.
The DMRG algorithm can be understood as a complete-active-space (CAS) CI method.
Compared to most other electron-correlation methods, much larger active spaces can be treated within the DMRG approach
without a predefined truncation of the complete $N$-particle Hilbert space. This feature of the
algorithm allows one to obtain qualitatively correct wave functions and energies for very difficult electronic structures
\cite{dmrg_chan,marti2010b}. 

The \emph{numerical} DMRG optimization avoids the application
of a predefined \emph{analytic} CI expansion of the wave function, and hence, 
no straightforward connection can be established between the DMRG basis states
and the linear expansion in terms of Slater determinants. This feature of the algorithm is actually the reason why DMRG can converge CASCI
wave functions for very large active spaces, for which
an explicit construction of the complete many-particle basis is prohibitive (for an example see Ref.\ \cite{chan_water}).
One might argue that the explicit knowledge of the CI expansion coefficients is not needed 
as the one- and two-particle density matrices produced by DMRG are sufficient to calculate 
physical observables. However, detailed information about the underlying wave function should
not be disregarded as it can be used for the following purposes.

\begin{enumerate}
\item 
An analysis of DMRG wave functions in terms of CI expansion coefficients will be useful to better
understand the quality of the converged DMRG result as the electronic energy is not sufficient as a sole
criterion to assess the accuracy obtained \cite{gerrit_sd}. Also for well-established electron-correlation models based on the CC approach
Hanrath~\cite{hanrath08,hanrath09} has observed for the H$_4$ and HF model systems that considering the energy as the only
convergence criterion is not sufficient. Only an analysis of the underlying CC wave function revealed 
artifacts in the wave function expansion. In other words, in CC calculations one may obtain a wave function expansion that differs
notably from an FCI reference although the corresponding CC energy seems to be well converged towards the FCI
reference energy. As a consequence, Hanrath and co-workers advocated for a comparison of wave functions rather than energies to assess the quality of
a calculation. The same holds true for DMRG wave functions.
\item 
Since the CI decomposition proposed can be obtained for any converged DMRG calculation,
different DMRG calculations can be compared. This is beyond a purely numerical comparison
of expectation values and allows one to assess whether qualitatively similar wave functions were
actually obtained after DMRG convergence, whose main disadvantage is its purely numerical procedure
with convergence properties that are difficult to control and to steer.
\item 
Accordingly, the CI decomposition may help to optimize DMRG convergence as one can study when and why
various electronic configurations are picked up.
Reconstructing analytic wave functions facilitates the understanding and analysis of
electronic structure in terms of coefficients of Slater determinants. This is hardly possible
from the matrix product states (MPS) \cite{oestlund95,oestlund97} which is optimized by the DMRG algorithm.
\item 
A fast DMRG calculation with a small number of renormalized system states, which is always feasible and routinely possible in terms of computer
time, may be used to sample important determinants which may then be used as input
to a multi-reference CI procedure.
\end{enumerate}

A determinantal analysis of DMRG states has already been presented by our group \cite{gerrit_sd}. 
The 'transcription' scheme of DMRG states into a
Slater determinant basis used in that work is, however, only applicable to a small number of active orbitals. Larger active spaces
containing $L$ spatial orbitals remain computationally unfeasible since {\it all} states in the
Fock space spanned by these $L$ orbitals were kept. The number of basis states grows, of course, exponentially
with the number of orbitals $L$ (e.g., as $4^L$ for spatial orbitals).
Thus, the fact that DMRG allows us to obtain results for very large active spaces well beyond the binomial
scaling wall of common multi-reference methods, i.e., beyond the binomially growing dimension of the
complete $N$-particle Hilbert space, which scales as $ {\displaystyle L \choose N_\alpha }\cdot {\displaystyle L \choose N_\beta }$
with ${\displaystyle L \choose N_\sigma }= \frac{\displaystyle L!}{\displaystyle N_\sigma ! (L-N_\sigma)!}$  
for $\sigma\in\{|\uparrow \rangle,|\downarrow \rangle\}$, has not been exploited yet.
As a consequence, an improved decomposition scheme is required, which can be employed for any dimension
of the active space.  So far, the direct calculation of CI coefficients from MPS \cite{oestlund95,oestlund97}
has not been explicitly carried out for molecular electronic structures and we will show here how this can be accomplished
for active spaces larger than those feasible for FCI-type approaches.

This work is organized as follows. In section\ \ref{sec:theory}, we present our
improved decomposition scheme
of the DMRG wave function. Then in section~\ref{sec:sampling}, we discuss the sampling routine to
explore the binomially large Hilbert space efficiently by means of a Monte
Carlo approach, which is then tested at the ozone transition state structure and applied to reconstruct
the CI coefficients for a large active space of the Arduengo carbene in
section \ref{sec:results}. Finally, a summary and concluding remarks are given
in section~\ref{sec:conclusion}.

\section{CI expansion of DMRG states \label{sec:theory}}

The total DMRG state $\Psi^{(s)}(N)$ describes $N$ electrons at a given DMRG (micro)iteration step
$s$ and is given by
\begin{equation}\label{dmrgwfu}
\Psi^{(s)}(N) = \sum_{ij} \psi_{ij}^{(s)} \left|i\right\rangle \otimes \left|j\right\rangle
\end{equation}
with expansion coefficients $\psi^{(s)}_{ij}$. \{$\left|i\right\rangle$\} and \{$\left|j\right\rangle$\} represent
the orthonormal product basis of the system (also called active subsystem) and the environment (complementary subsystem), respectively. 
These basis states are
never explicitly constructed \cite{mitr03,gerrit_sd}, only the expansion coefficients $\psi^{(s)}_{ij}$ are obtained in each step
of the algorithm. We assume a two-site DMRG protocol.

A common expansion of total wave functions employed in quantum chemistry consists of a linear combination
of Slater determinants $\Phi_{\{\bf n \}} = | n_1 \ldots  n_{L}\rangle$. Here, $\Phi_{\{\bf n \}}$ is given
in its occupation number representation and $\{\bf n \}$ represents the set of all occupation number vectors
constructed from $L$ one-particle states. The total DMRG wave function may then be written 
as a CI-type wave function, 
\begin{equation}\label{slaterwfu}
\Psi^{(s)}(N)=\sum_{\{\bf n \}} C_{\{\bf n\}}^{(s)}  \Phi_{\{\bf n \}},
\end{equation} 
where the expansion coefficients $C_{\{\bf n\}}^{(s)}$ are the CI coefficients that are to be determined. 
The sum in Eq.\ (\ref{slaterwfu}) runs over all occupation number vectors in the $N$-particle Hilbert
space with the correct particle number, projected spin, and point-group symmetry.

It is important to understand that the two sets of expansion coefficients, $\{\psi_{ij}^{(s)}\}$ and $\{C_{\{\bf n\}}^{(s)}\}$, are not identical,
but the CI coefficients $\{C_{\{\bf n\}}^{(s)}\}$ can be reconstructed from the DMRG expansion coefficients $\{\psi_{ij}^{(s)}\}$
and the renormalization matrices \cite{gerrit_sd}.
For the explicit reconstruction of CI coefficients for a DMRG wave function, 
the work of \"Ostlund and Rommer \cite{oestlund95,oestlund97} showed that the DMRG algorithm
establishes a recursion relation for matrix product states. They considered projection operators
$\hat{A}_i[n_i]$ dependent on the local site $n_i$ which map from one $m$-dimensional space
spanned by an orthonormal basis \{$\left| m_{l-1} \right\rangle$\} to another $m$-dimensional space
spanned by \{$\left| m_l \right\rangle$\} where $l$ denotes the number of molecular orbitals already incorporated in the active
subsystem. These projection operators can be represented by
$m\times m$ matrices $({A}_i[n_i])_{m_l;m_{l-1}}$. In the renormalization step of the DMRG algorithm
the dimension of the product basis of the enlarged system block \{$\left|m_{l-1}\right\rangle
\otimes \left|n_l\right\rangle$\} is reduced to a set of
new basis states which represent the most important states \{$\left|m_l\right\rangle$\} of the enlarged
block,
\begin{equation}
\left| m_{l}\right\rangle = \sum_{m_{l-1}, n_{l} } (A_{l}[n_l])_{m_{l};{m_{l-1}}} |m_{l-1} \rangle \otimes
|n_{l}\rangle,
\end{equation}
where the tensor product $\textstyle|m_{l-1} \rangle \otimes |n_{l}\rangle$ is a basis state of the active subsystem.
We can now perform a backward recursion to
express $\left|m_{l-1}\right\rangle$ through $\left|m_{l-2}\right\rangle$ and so forth, until we reach
the first site of the lattice. For a system block of length $l$ we then obtain
\begin{equation}\label{ml-expansion}
\left| m_{l}\right\rangle = \sum_{n_1 \ldots n_l}\left(A_{l}[n_l] \ldots A_{2}[n_2]
\right)_{m_l; n_1} | n_1  \ldots  n_{l}\rangle,
\end{equation}
with $\left| n_1  \ldots  n_{l}\right\rangle = \left| n_1 \right\rangle \otimes \left| n_2 \right\rangle
\otimes \ldots \otimes \left| n_l\right\rangle$. The DMRG states of the environment block can be expanded
in a similar way. Combining Eqs.\ (\ref{ml-expansion}) and (\ref{dmrgwfu}) and the corresponding recursion
relation for the environment states, we can write for the superblock wave function 
\begin{align}
\Psi^{(s)}(N) = &\sum_{\{\bf n\}} \sum_{m^S} \sum_{m^E}
\psi_{m^S n_{l+1} n_{l+2}m^E}^{(s)} \label{dum}\\ \nonumber
 &\times \left(A_{l}[n_l] \ldots A_{2}[n_2] \right)_{m^S; n_1} 
\times \left(A_{l+3}[n_{l+3}] \ldots A_{L-1}[n_{L-1}]\right)_{m^E; n_L} 
\times \left| n_1  \ldots  n_{L}\right\rangle.
\end{align}
for a lattice of length $L$.
Hence, the superblock wave function lives in a space spanned by two matrix product states and a product of two
local sites \cite{schollwoeck,dmrg_chan,marti2010pccp}. 
By comparing Eqs.\ (\ref{slaterwfu}) and (\ref{dum}), we can now identify the CI coefficients for the $s$-th microiteration step
\begin{eqnarray}
\label{ci_coeff_eq}
C_{\{ \bf n \}}^{(s)} &=  &\sum_{m^S}^m \sum_{m^E}^m \psi_{m^S n_{l+1} n_{l+2}m^E}^{(s)} \\ \nonumber 
&&\times \left(A_{l}[n_l] \ldots A_{2}[n_2] \right)_{m^S; n_1} 
\times \left(A_{l+3}[n_{l+3}] \ldots A_{L-1}[n_{L-1}]\right)_{m^E; n_L}.
\end{eqnarray}
The CI coefficients can be determined for any composition of the lattice blocks as soon as all lattice-position-dependent
transformation matrices $A_{j}$ are obtained which are optimized during one sweep of the DMRG algorithm.
Note that these transformation matrices are represented as $4m\times m$ matrices
in the DMRG algorithm and need to be decomposed into four $m\times m$ matrices
for each possible occupation of site $j$ ($|\uparrow\downarrow\rangle,
|\downarrow \rangle,|\uparrow  \rangle,|  \rangle$).

\section{Exploring the binomially large many-particle Hilbert space \label{sec:sampling}}
In the previous section, we have discussed how to calculate CI coefficients for
a given occupation number vector within the DMRG framework. 
It is, however, unfeasible and unnecessary to create the entire basis of the $N$-particle
Hilbert space since the numerical evidence compiled over the years
demonstrates that only a fraction of the entire set of occupation number vectors is
decisive for a sufficiently reliable representation of the wave function 
\cite{knowles,knowles_fci,greer98,Mitrushenkov,feller,greer}. In the language of DMRG, this
translates to our previous observations \cite{marti2008,marti2010b} that
already small-$m$ DMRG calculations provide accurate and qualitatively correct
relative energies between two spin states or two isomers on the same potential
energy surface. 
Note that small-$m$ DMRG calculations can, in principle, also describe
non-sparse determinantal wave functions.
Actually, coupled cluster and
truncated configuration-interaction theories also heavily rely on the fact that
determinants following only a certain excitation pattern (i.e., orbital substitution pattern)
are included in the wave function and still provide very accurate electronic energies
for many molecular systems.

\subsection{The SR-CAS algorithm}
Because of the formally binomially scaling of many-particle states (determinants) in the $N$-particle Hilbert space,
an algorithm is required that is capable of sampling important determinants.
The basic idea of the sampling routine is related (but not restricted) to the underlying MPS
because the DMRG algorithm locally optimizes the parameter space over the
MPS~\cite{marti2010b}.
Hence, a local update in an occupation number vector with a large
weight might lead more likely to another occupation number vector with another large
weight.

Our sampling routine --- inspired by our previous work on the variational
optimization of tensor network states \cite{cgtnarxiv2010} and by the work of Sandvik and
Vidal~\cite{sandvik2007} --- performs 
an excitation of the type 
$a_{p_1}^\dagger 
a_{q_1}
a_{p_2}^\dagger 
a_{q_2}
\dots
a_{p_i}^\dagger 
a_{q_i}$
from a predefined reference (usually the Hartree--Fock determinant), 
where $i$ is a random number between $1$ and $N$ ($p_j$, $q_j$ are also random
numbers in the interval from $1$ to $L$), preserving the number of
particles, projected spin, and point-group symmetry. 
This procedure is capable of sampling the complete FCI space. 
Each step in our sampling procedure consists of three
steps. In the first step, we draw a random number between $1$ and $N$ to determine how many
excitations --- each represented by an excitation operator $a_{p_j}^\dagger
a_{q_j}$ --- are performed from a reference determinant (still choosing the Hartree--Fock determinant
at the start). For each of these excitations, we generate in the second step  two further
random numbers within the interval of $1$ and $L$ to determine the spin orbitals
of the reference that is to be exchanged by one from the virtual space. In the third step, all excitation
operators are applied to the reference determinant which then yields a new trial determinant (occupation
number vector).

The coefficient of the trial occupation number vector is then calculated
according to Eq.\ (\ref{ci_coeff_eq}) and stored in a hash data structure if it is above a predefined
threshold. 
As we are not interested in all tiny coefficients, we store only the relevant basis states with a
coefficient larger than the predefined threshold.
The threshold value also prevents a binomial scaling in the dimension of the hash.
We have employed a thread-safe hash data structure that allows a massively
parallel execution of the sampling routine.
In the following we abbreviate the sampling--reconstruction algorithm for FCI-type wave function defined
in a complete active orbital space as SR-CAS.

\subsection{Completeness measure for accuracy control}
In order to extend the search over the entire Hilbert space, 
we update the reference occupation
number vector $\Phi_{\bf n}$ by a new one $\Phi_{\bf m}$ 
according to the probability
\begin{equation}
p( \Phi_{\bf n} \leftrightarrow  \Phi_{\bf m}) =
	\min \left( 1,  \frac{ | C_{\bf m} |^2 }{ | C_{\bf n} |^2 }
			\right).
\label{prob}
\end{equation}

The accuracy and quality of the reconstructed CI-type wave function can be monitored by means
of the sum over the squared CI coefficients of the already considered occupation number
vectors. For this we define a completeness measure (COM),
\begin{equation}
{\rm COM } = 1 - \sum_{\{\bf n\}} | C_{\{ \bf n \}} |^2
\label{tre}
\end{equation}
which quantifies the amount of CI coefficients of occupation number vectors that have not yet been sampled.
It must be noted that the COM defines a convenient convergence criterion for the sampling routine
as it must converge to zero upon sampling of the full determinant space (covered by the DMRG wave function). As a consequence, it
can be employed to estimate the quality of the explicit CI-type wave function obtained and of the 
expectation values calculated with this wave function. In contrast to common threshold measurements defined
for the energy as for instance applied by Buenker and Peyerimhoff \cite{peyerimhoff74,peyerimhoff75,peyerimhoff77} in multi-reference CI calculations,
COM directly refers to the quality of the reconstructed CI wave function since the reference wave function is known and orthonormal.
We should note that our SR-CAS algorithm is considered converged when the actual COM value calculated for the reconstructed
CI expansion falls below a predefined threshold. Often we use the same threshold as the one needed
in the sampling routine described above, except in cases where it turns out that smaller CI coefficients are important
for the predefined threshold of COM.

\subsection{Quantum fidelity as a measure for DMRG convergence}

A measure to assess the convergence of the DMRG algorithm 
is highly desirable~\cite{marti2010a,koni_diss}. Usually, DMRG studies perform several calculations
with an increasing number of DMRG system states to verify the convergence of the
algorithm. The DMRG energy is then usually taken as the sole convergence criterion. There
exist other approaches to execute a DMRG calculation which are 
based on a dynamic selection of the DMRG system states
introduced by Legeza \emph{et al.} \cite{lege03a,lege04}. But even his dynamic block
state selection (DBSS) procedure requires several calculations to be
carried out as the convergence is also monitored by means of the energy
expectation value. 

With our SR-CAS algorithm at hand, we can now
examine a measure of convergence for a sequence of DMRG calculations that
are either performed by increasing the number of DMRG block states $m$ or by the
DBSS procedure. This measure is based on the likeness --- the \emph{fidelity} ---
of two quantum states~\cite{peres1984,zhou2008}. 
Already in 2003, Legeza and S{\'o}lyom mentioned the potential application of the fidelity concept
within the DMRG framework~\cite{lege03c}. The fidelity $F$ between two converged DMRG states is
defined as
\begin{equation}
F_{m_1, m_2} = \vert{\ \langle{\Psi_{m_1}^{(s)}}}\vert{ \Psi_{m_2}^{(s)}\rangle{}}\ \vert{}^2
\label{fidelity}
\end{equation}
where $s$ stands for a given microiteration, e.g., for a specific partitioning
of the orbital space among the active and complementary subsystems, $\Psi_{m_1}$ and $\Psi_{m_2}$ denote two
DMRG states obtained for different numbers of DMRG block states, $m_1$
and $m_2$, respectively.
Qualitatively speaking, the quantum fidelity measure expresses the overlap between the two states.
It can thus also detect a sign change of the CI
coefficients which might not significantly affect the energy at all.

The value of the quantum fidelity measure is obvious for cases in which the energy turns out to be
no reliable convergence criterion, because the underlying wave functions
differ (as can then be detected by the fidelity $F$). In our previous DMRG study on the
decomposition of DMRG basis states into a Slater determinant basis~\cite{gerrit_sd}, 
we had already observed that the energy can decrease during DMRG iterations, while the
corresponding DMRG wave function was qualitatively wrong when compared to the FCI solution.
We should recall from the introduction that similar results have been reported in the
literature for CC calculations~\cite{hanrath08,hanrath09}.

It is now important to note that Eq.\,(\ref{fidelity}) cannot be straightforwardly evaluated 
because the DMRG basis states are different for the $m_1$
and $m_2$ calculations. Hence, a simple scalar product of the two expansion
coefficient vectors does not yield the overlap measure. This is solved in our work here.
The CASCI-type wave function, which we can now generate with our SR-CAS
scheme, allows us to calculate the quantum fidelity defined in Eq.\,(\ref{fidelity}).

\section{A multireference test molecule\label{sec:ozon}}

In order to study the performance of our reconstruction procedure, we reconsider the 
multireference molecule that has already served as a test case in our original determinant decomposition
of DMRG states \cite{gerrit_sd} (the structure was taken from Ref.\ \cite{ozonstructure}
and features one short O--O bond length
of 121.1 pm, one long O$\cdots$O bond distance of 240.7 pm and an OOO bond angle of 51.1$^\circ$).
The calculation of integrals in the molecular orbital basis was performed in $C_{2v}$ point group symmetry with the \textsc{Molpro} program package \cite{molpro}
employing Dunning's cc-pVTZ basis set \cite{dunning,dunning2}. The DMRG calculations were carried out with our 
program \cite{dmrg_new}.
As active space, eight electrons correlated in nine orbitals were chosen. Hence, the complete
many-particle Hilbert space is spanned by 7'956 Slater determinants.

The transition state energies obtained for different quantum chemical methods are shown in Table \ref{energiesozon}.
For increasing $m$, the DMRG transition state energy converges quite fast towards the CASCI reference which
is then reproduced for $m=156$ DMRG system states.
\begin{table}[h]
\caption{Transition state energies for ozone in a cc-pVTZ 
basis set for different quantum chemical methods.
The active space comprises 8 electrons correlated in 9 orbitals.
The Hartree--Fock and CASCI results were taken from Ref.\ \cite{gerrit_sd}.
\label{energiesozon}
}
{
\begin{center}
\begin{tabular}{lc}\hline \hline
Method & E/Hartree \\\hline
HF & $-$224.282 241 \\ 
CASCI & $-$224.384 301 \\ \hline
DMRG($m=48$) & $-$224.384 216 \\
DMRG($m=100$) & $-$224.384 300 \\
DMRG($m=156$) & $-$224.384 301
\\ \hline
\hline
\end{tabular}
\end{center}
}
\end{table}

Since the dimension of the many-particle Hilbert space spanned by these nine orbitals is rather small (only 7'956
Slater determinants), it is possible to determine the CI coefficients for all Slater determinants
directly with the \textsc{Molpro} program package \cite{molpro}. This is advantageous as it allows us to
test the efficiency of our SR-CAS Monte Carlo sampling routine.
As described before, only Slater determinants are accepted with CI coefficients larger than a predefined threshold.
A large threshold implies, of course, that quite many 'unimportant' configurations will be rejected.
In Fig.\ \ref{fig:ozon_energy}, the transition state energy corresponding to the CASCI-type expansion composed of
all sampled configurations is shown for a different number of DMRG system states $m$. As one would expect, for decreasing threshold
the approximate transition state energy converges towards the CASCI reference energy for all
values of system states $m$. It is obvious that the approximate transition state energy is the lower the larger $m$ is 
(see, e.g., the inlay in Fig.\,\ref{fig:ozon_energy}).
A threshold of $0.0001$, which results from 1'532 explicitly considered determinants,
is sufficient to obtain an accurate CASCI-type wave function expansion for which the
transition state energy deviates by 1.371 mHartree (3.6 kJ/mol) from the CASCI reference energy for $m=156$.
Whether an energy difference of this order can be tolerated or not depends on the accuracy that is desired.
For many relative energies sought in chemistry, like reaction energies in transition metal catalysis 
\cite{markus_chimia_2009,marti2010pccp,markus_fd2} or excited states \cite{dreu2005,dreu2006,neug2010,neug2011}, this would be sufficiently accurate.
A further decrease by two orders of magnitude to a threshold of $10^{-6}$, which now includes 5'650 explicitly considered determinants,
leads to a deviation by 11 $\mu$Hartree (0.029 kJ/mol) from the CASCI reference.

\begin{figure}[h]
\centering
\includegraphics[width=0.9\linewidth]{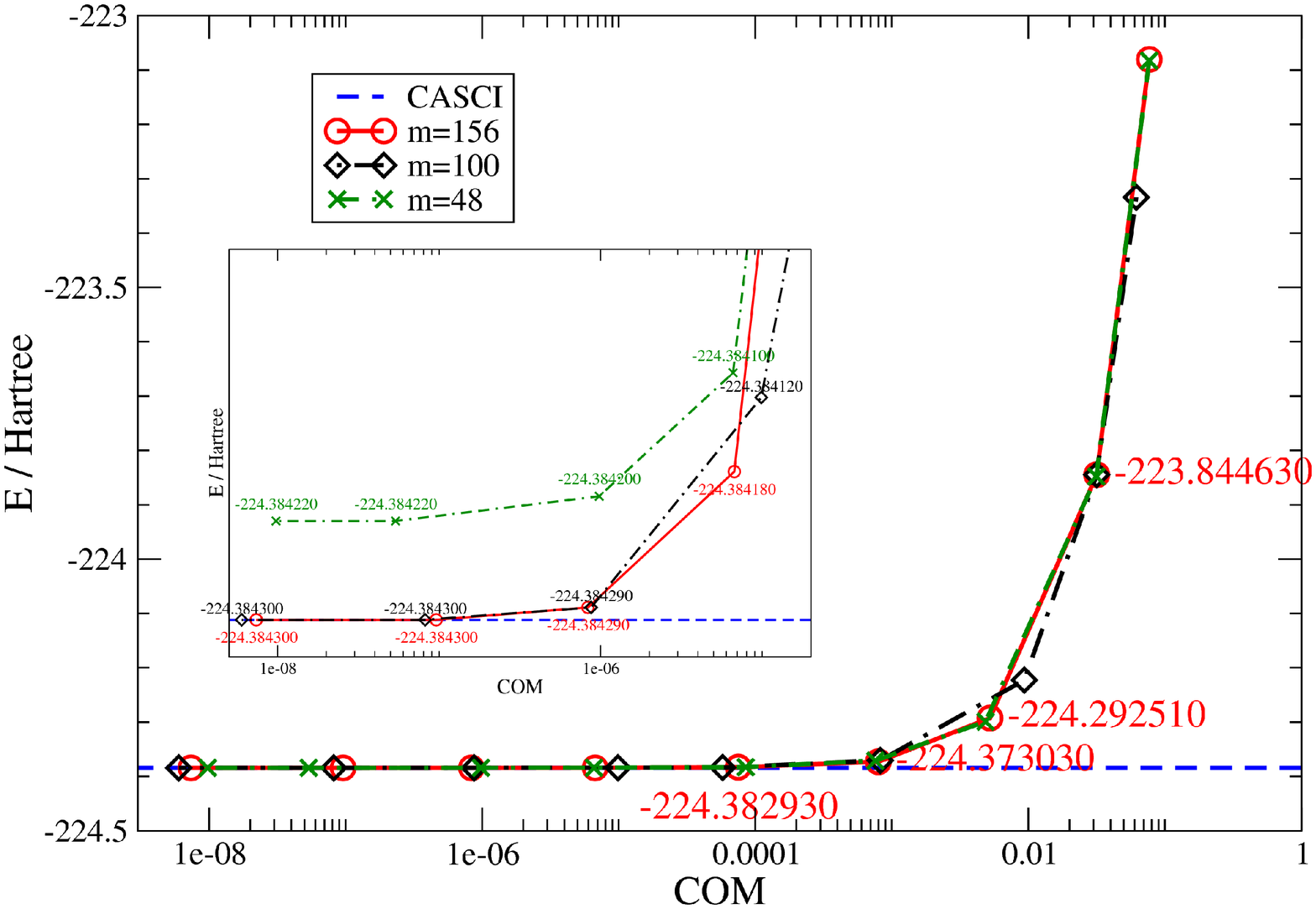}
\caption{Energy of the ozone molecule in its transition state structure calculated from sampled configurations
for different COM values in Hartree. The dashed blue line represents the CASCI reference energy taken from Ref.\ \cite{gerrit_sd}.
}\label{fig:ozon_energy}
\end{figure}

The number of Slater determinants which span the sampled subspace of the complete many-particle Hilbert space
is shown in Fig.\,\ref{fig:ozon_sd}. In general, more configurations are needed
for larger $m$-calculations in order to reach the desired accuracy, i.e., to fall below a specific threshold, as
compared to smaller $m$-calculations. For $m=48$, a kink in the graph can be observed that is not present for
larger $m$, which indicates that the DMRG calculation has not incorporated all relevant configurations. 
Because of these missing configurations, the DMRG algorithm converges to a higher energy for $m$=48
(note that this failure is cured by the DBSS procedure \cite{lege03a,lege04}).

As mentioned above, an accurate CASCI-type wave function expansion for $m=156$ is already obtained for a
threshold of $0.0001$ which can be expanded in only 1'532 Slater determinants (20\% of the complete many-particle Hilbert space).
All these Slater determinants correspond to CI coefficients larger than $10^{-4}$. The large
percentage of the Hilbert space needed is easy to understand in view of the small active space,
which limits the total number of Slater determinants that can be constructed. From a 
statistical point of view, the ratio of relevant configurations will decrease with increasing dimension of the
Hilbert space (i.e., with increasing size of the CAS).

To obtain highly accurate energies in the $\mu$Hartree regime, the threshold can be reduced to
$10^{-6}$, where all important configurations are already sampled spanning a subspace
consisting of 5'078 Slater determinants (64\% of the complete many-particle Hilbert space).
The consideration of further 'unimportant' configurations does not change the corresponding
transition state energy (see again inlay in Fig.\,\ref{fig:ozon_energy}). 
Note that an accuracy of about 1 kJ/mol in the total electronic energy would require a COM value of
about $10^{-5}$.
The flattening of the sampling curve when approaching COM $=0$
indicates that only a few configurations are left which have non-zero CI coefficients 
(but still are much smaller than $10^{-6}$). 
Hence, the DMRG optimization only picks up the most important configurations
and the many-particle Hilbert space is being truncated for $m=48$.

\begin{figure}[h]
\centering
\includegraphics[width=0.7\linewidth]{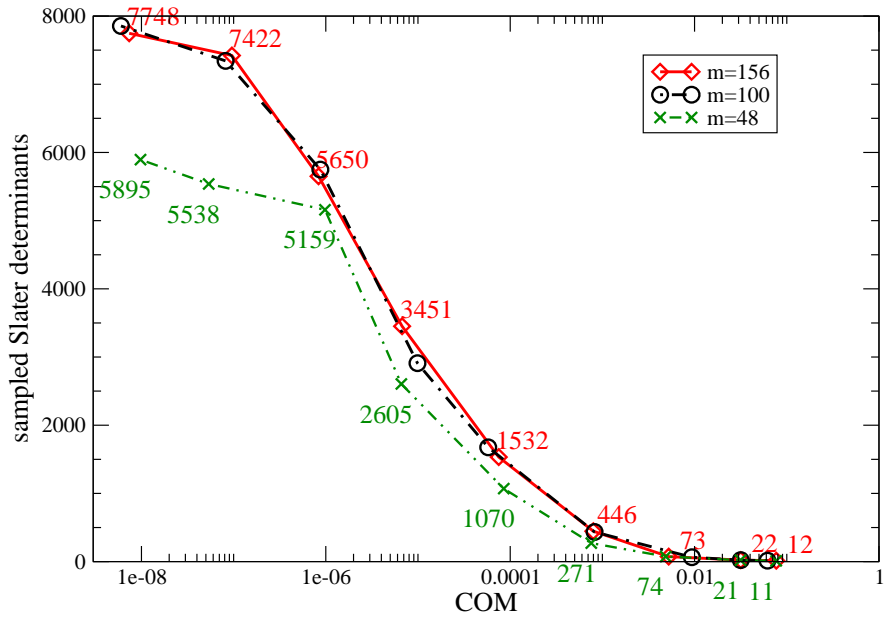}
\caption{Number of sampled Slater determinants for DMRG calculation employing different $m$ values
for the transition state structure of ozone.
The total Hilbert space has a dimension of $7'956$.
A sufficiently accurate wave function can already be obtained for
a COM of about $10^{-4}$ which corresponds to 1'532 Slater determinants (about 20\% of the complete
many-particle Hilbert space.)
}\label{fig:ozon_sd}
\end{figure}

A reconstructed analytic CASCI-type wave function can provide further insights into the accuracy
of DMRG. Legeza and coworkers \cite{legeza_molphys} speculated that if the accuracy of DMRG calculations is lowered,
i.e., if $m$ is decreased, the structure of the wave function will essentially be preserved.
This assumption was based on the expectation values, whereas a direct comparison of the wave functions was not
performed. With the scheme proposed here, we can now explicitly analyze the dependence of the structure of the CASCI-type wave
function expansion on the number of DMRG system states $m$. This can be accomplished by examining the excitation
patterns with respect to the configuration with largest CI coefficient for different threshold values and $m$ values 
which are shown in Fig.\,\ref{fig:ozon_ex}. Generally, if
$m$ is increased, more configurations corresponding to higher excitations are included in the wave function expansion
at constant threshold. Exceptions can be observed if more configurations with small CI coefficients are
sampled so that in total more Slater determinants are needed to reach the desired threshold for COM. Moreover, by systematically
lowering the threshold for the consideration of CI coefficients in the sampling routine, 
configurations containing higher excitations are picked up (up to octuple
excitations) and the maximum of the
excitation pattern is shifted towards higher excitations. 

The total excitation patterns are similar for all $m$-calculations.
We should emphasize that even for $m=48$ octuple excitations are present in the DMRG wave function.
In this case with $m=48$, the excitation pattern does not change considerably for a threshold $\leq 10^{-6}$ (see also Fig.\,\ref{fig:ozon_sd})
and only a few highly excited Slater determinants are picked up. Important higher excitations are missing
(too small or zero CI coefficients) in the wave function and the number of DMRG system states $m$ needs to be
increased to consider them appropriately. For the fully converged CASCI-type wave function
(COM $\leq 10^{-7}$), we observe qualitatively similar excitation patterns for a different number of DMRG
system states $m$. In principle, any $n$-fold excitation (heptuple, octuple, ...) is accessible by the DMRG algorithm.
Most importantly, the total
structure of the DMRG wave function is basically retained for different thresholds of accuracy.

\begin{figure}[h]
\centering
\includegraphics[width=0.9\linewidth]{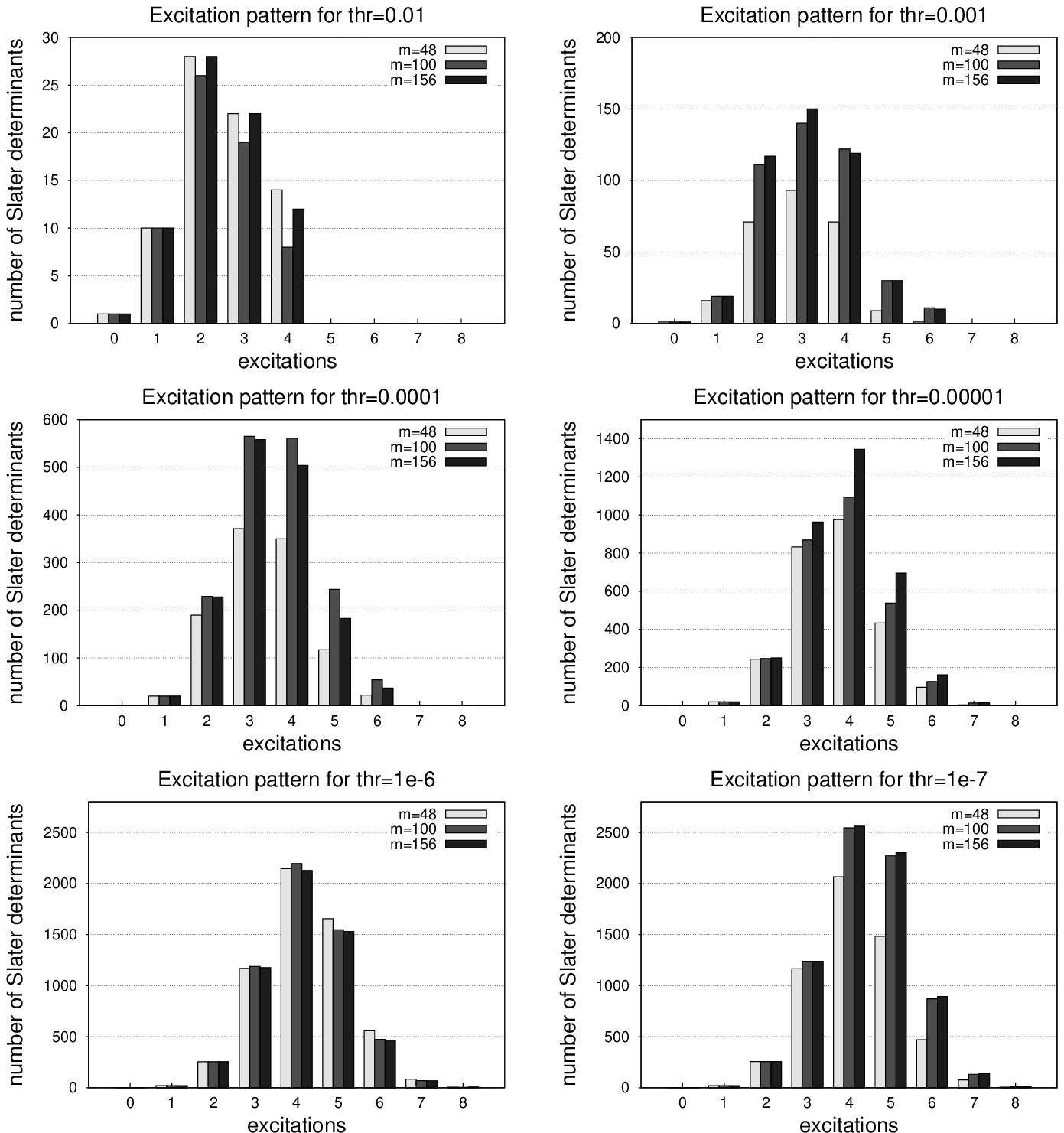}
\caption{Comparison of excitation patterns with respect to the configuration 110110000000111001 for DMRG
calculations on the ozone molecule employing different $m$ values and with respect to a chosen threshold (thr) for the consideration of 
CI coefficients in the sampling routine. Note the individual
scaling of the ordinate for an optimal representation of the excitation spectrum.
}\label{fig:ozon_ex}
\end{figure}

Fig.\ \ref{fig:ozon_ex_156} shows the excitation pattern for the DMRG($m=156$) calculation depending 
on the value chosen for threshold in greater detail. As already mentioned, decreasing the threshold incorporates an increasing number of 
Slater determinants with a small CI coefficient
in the CASCI-type wave function and shifts the maximum of the excitation pattern towards higher excitations.
This is due to the fact that higher excitations have smaller CI coefficients and are only
sampled if the threshold value is small enough. Note also that
our SR-CAS algorithm is capable of picking up the most important configurations first,
which correspond to single and double excitations and some triple excitations. 
To obtain a highly accurate CASCI-type wave function expansion
(COM $\leq 0.00001$), 
higher excitations (mainly quadruple, quintuple and hextuple) must be included as well.

\begin{figure}[h]
\centering
\includegraphics[width=0.6\linewidth]{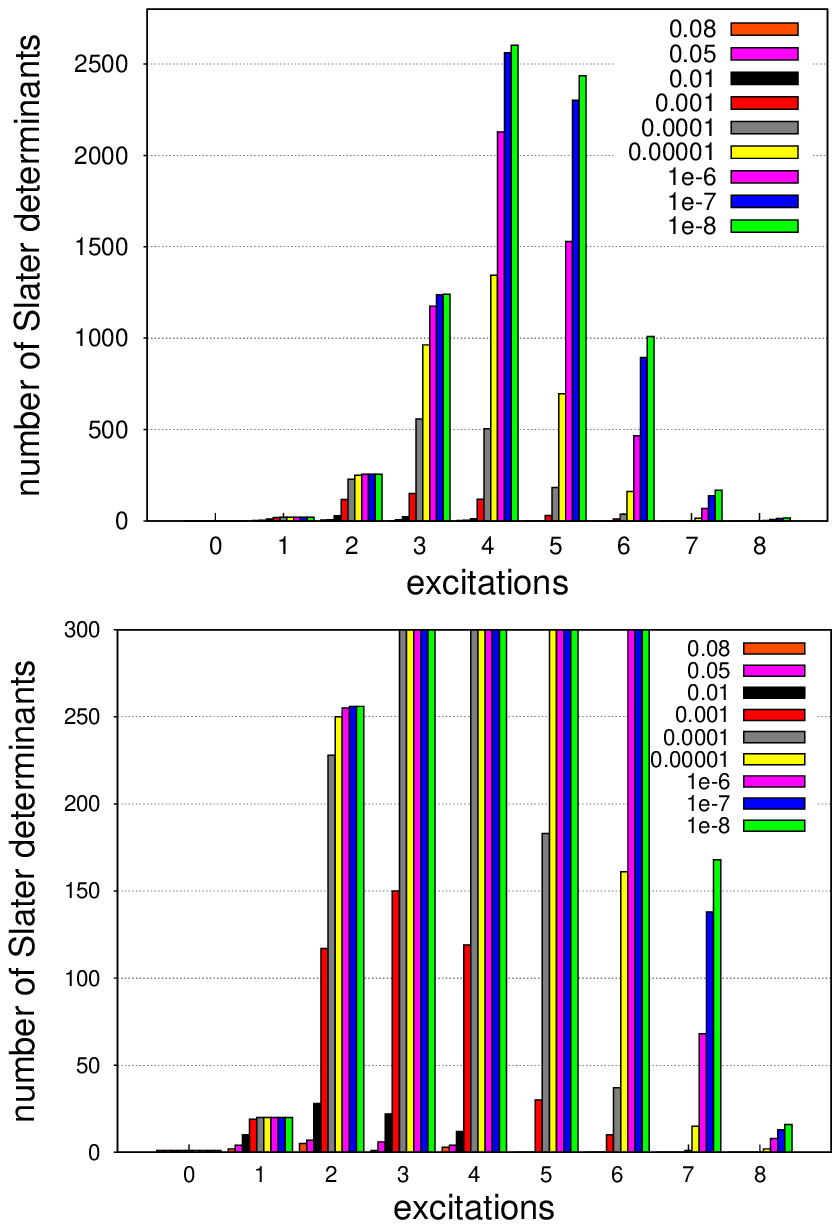}
\caption{Excitation patterns for the DMRG($m=156$) calculation on the transition state structure of ozone
employing different threshold values for the consideration of CI coefficients in the sampling routine. Excitations are
given with respect to the 110110000000111001 configuration as reference. 
The upper graph shows the total excitations spectrum, while the lower half represents a cutout
to emphasize the lower and higher excitations, respectively.
}\label{fig:ozon_ex_156}
\end{figure}

The most important CI coefficients are shown in Table \ref{tab:ciozon}. For an increasing number of DMRG system
states $m$, the changes in CI coefficients are small. 
As the energy still changes notably, this implies that small changes in CI coefficients lead to larger changes in the energy. 
For $m=156$, the corresponding CI coefficients are in good agreement with
the CASCI reference calculation \cite{gerrit_sd}.

\begin{table}[h]
\caption{CI coefficients of selected configurations of ozone in a transition state structure.
The definition of the active space and the labelling of the molecular orbitals were taken from Ref.\ \cite{gerrit_sd}.
Doubly occupied orbitals are denoted as '2', empty orbitals as '0', and orbitals singly occupied by a spin-up or
spin-down electron as '+' and '$-$', respectively.
\label{tab:ciozon}}
{
\begin{center}
\begin{tabular}{ccccccccccccc}\hline \hline
\multicolumn{9}{c}{Occupation} & \multicolumn{3}{c}{DMRG} & \multirow{2}{*}{CASCI} \\ \cline{1-9}
$9a^\prime$&$10a^\prime$&$11a^\prime$&$12a^\prime$&$13a^\prime$&$14a^\prime$&$1a^{
\prime\prime}$&$2a^{\prime\prime}$&$3a^{\prime\prime}$& $m=48$     &   $m=100$     &   $m=156$    & \\\hline
2&+&--&0&0&0&2&--&+  &    0.437 359  &  0.437 246   &    0.437 246 & 0.437 253 \\
2&--&+&0&0&0&2&+&--  &    0.437 359  &  0.437 246   &    0.437 246 & 0.437 253 \\
2&2&0&0&0&0&2&2&0    &    0.391 038  &  0.391 875   &    0.391 876 & 0.391 876 \\
2&2&0&0&0&0&2&+&--   &    0.287 457  &  0.287 161   &    0.287 160 & 0.287 160 \\
2&2&0&0&0&0&2&--&+   &    0.287 457  &  0.287 161   &    0.287 160 & 0.287 160 \\
2&--&--&0&0&0&2&+&+  &    0.175 057  &  0.175 412   &    0.175 412 & 0.175 405 \\
2&+&+&0&0&0&2&--&--  &    0.175 032  &  0.175 412   &    0.175 412 & 0.175 405 \\ \hline
\hline
\end{tabular}
\end{center}
}
\end{table}

To obtain a CASCI-type wave function that yields an accuracy for the energy expectation value of less than 1 kJ/mol 
compared to the reference energy, a COM of about $10^{-5}$ for the determinant selection in the SR-CAS algorithm is sufficient. Due to the predefined parameters of
our SR-CAS algorithm, the reconstructed wave function is a
sum over configurations with weights above $10^{-5}$. To demonstrate
that this is indeed the case, the relative distribution of CI
coefficients with respect to the dimension of the total Hilbert space is
plotted in Figure \ref{fig:figureXXX}. One can observe that the main part of the CASCI
wave function consists of configurations with smaller weights ($<10^{-4}$) and that our SR-CAS algorithm is
capable of picking up the
configurations with largest weights since almost all Slater determinants
with CI coefficients larger than $10^{-4}$ have been sampled (compare
Figure \ref{fig:figureXXX} which shows the relative distribution of CI coefficients for
the total Hilbert space).

\begin{figure}[h]
\centering
\includegraphics[width=0.9\linewidth]{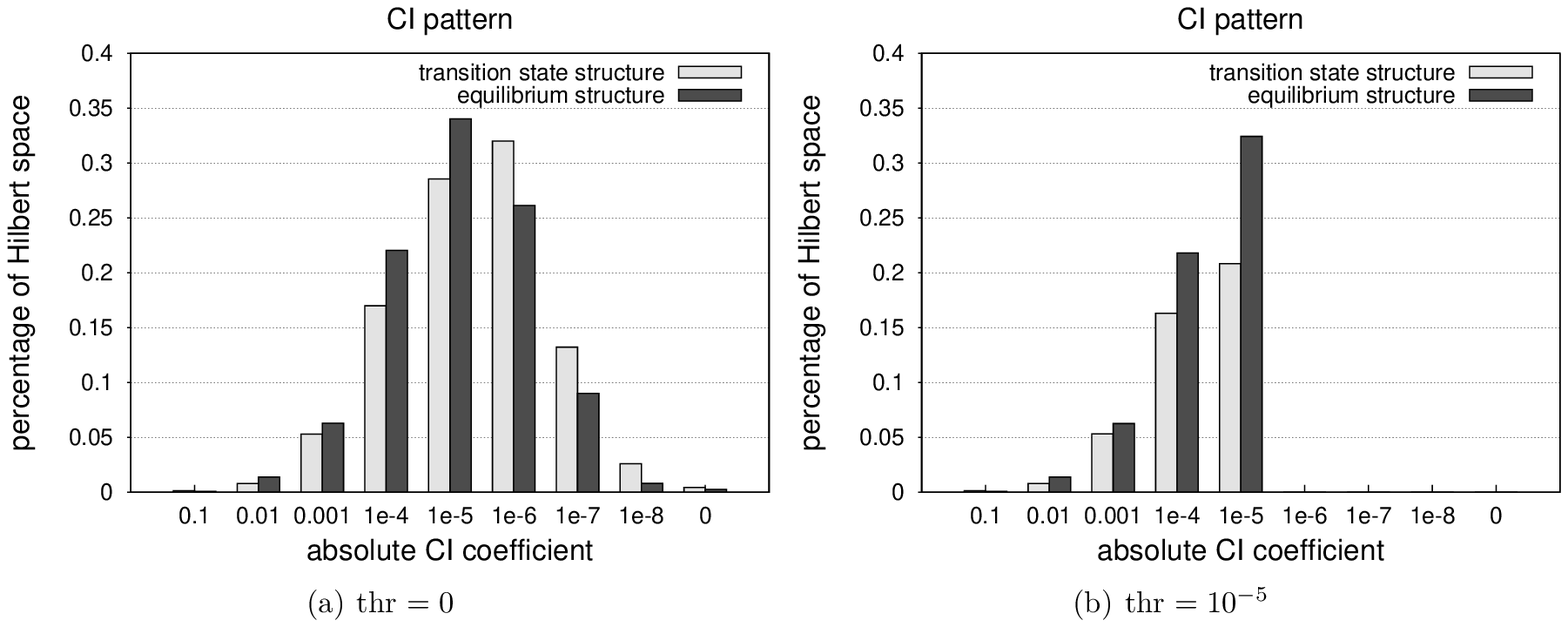}
\caption{Composition of the total Hilbert space measured in terms of the absolute values of CI coefficients
for ozone in its transition state and equilibrium structures. The
abscissa refers to configurations with CI weights between the given
value (lower bound) and the next higher value on the left (upper bound).
The ordinate gives the portion of the configurations considered
with respect to the total Hilbert space. 
(a) Structure of the total Hilbert
space. (b) Structure of the sampled configurations for a threshold of $10^{-5}$.
}\label{fig:figureXXX}
\end{figure}

For the multireference case, i.e., for ozone in its transition state structure,
our analysis provides an overview on the composition of the Hilbert space 
in terms of CI coefficient distribution. It is interesting to see
how the composition of the Hilbert space and its sampled
subspaces change when single-reference systems are considered. The ozone
molecule in its equilibrium structure represents such a typical
single-reference case.
In Figure \ref{fig:figureXXX}, the distribution
of (the absolute values of) CI coefficients for ozone in its transition state are compared to those of the
equilibrium structure. The total wave function of the single-reference
case is composed of more configurations with larger CI weights ($>
10^{-5}$), while for the multi-reference molecule, the maximum of the
curve is shifted towards lower-weighted configurations ($<10^{-6}$) than
it is observed for the single-reference case. This distribution of CI
weights indicates that for single-reference molecules a larger
percentage of the Hilbert space needs to be sampled for a given accuracy of the electronic
energy and thus to fall below a predefined threshold for COM. To
obtain a highly converged wave function (COM $\leq 10^{-5}$), considerably
more configurations out of the total Hilbert space are necessary for the
ozone molecule in its equilibrium structure than in its transition state
structure. Hence, our SR-CAS algorithm has to pick up a larger portion of
the Hilbert space if electronic structures with single-reference
character are considered. For this reason, we may well choose a single-reference case
to study our SR-CAS sampling-reconstruction algorithm as its reliability will strongly depend
on its capability to sample the relevant portion of the Hilbert space.

To conduct a proof-of-principle analysis of the fidelity measure at
the example of the ozone molecule, we calculate  for our three DMRG calculations with
$m=48,100,156$ the quantum fidelity. The set of these overlap measures is \{$(0.999'977'5)^2,(-0.999'999'9)^2$\} which indicates
that the changes in the wave function are small for increasing $m$. Increasing $m$ from $m=100$ to
$m=156$ DMRG block states corresponds to $F_{100,156}=(-0.999'999'9)^2$ which clearly shows the similarity
of both DMRG wave functions. Note that the minus sign implies a global sign change in the CI coefficient structure
which, of course, does not affect the structure of the wave function since the total wave function
can always be multiplied by a factor of $-1$.

\section{Reconstructed CI expansion for the Arduengo carbene \label{sec:results}}

In this section, we now study an active space that cannot be treated exactly because of its size.
We choose the 1,3-dimethyl Arduengo carbene 
(see Fig.\ \ref{fig:structure}) that was first experimentally characterized in 1992 \cite{arduengo_methyl}.
Although this molecule is a typical representative of a single-reference molecule and standard coupled-cluster
and even M{\o}ller--Plesset calculations may yield lower electronic energies,
such small aromatic molecules turn out to be rather difficult when their
excitation spectra shall be calculated with standard CASSCF-type methods because of the mixture of substituent
orbitals with orbitals from the aromatic ring (compared to the unsubstituted imidazole-2-ylidene). 
As a consequence, the methyl substituents lead to an active space of 16 electrons to be correlated in 28 molecular orbitals
in a CASCI- or CASSCF-type approach. 
Although we do not aim at specific properties of the Arduengo carbene in this work, it
is a suitable test case to make the point that we can sample a CAS that is not feasible to study
in a CASCI (i.e., FCI) or CASSCF approach. 

\begin{figure}[h]
\centering
\includegraphics[width=0.4\linewidth]{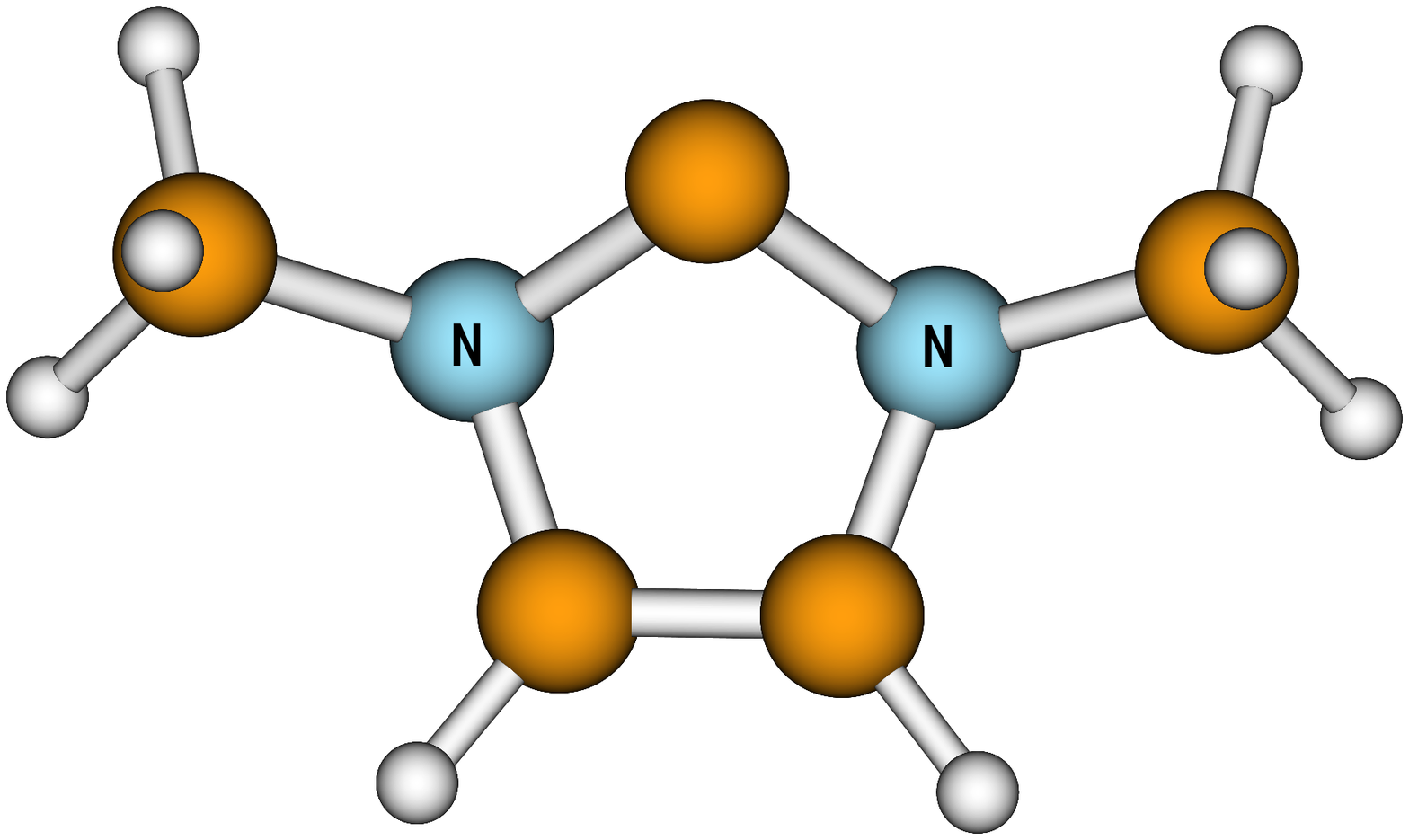}
\caption{BP86/TZP \cite{p86,b88}
optimized structure of 1,3-dimethyl imidazole-2-ylidene in its triplet state. The optimization was
performed using the quantum chemical program package \textsc{ADF} \cite{adf}.
}\label{fig:structure}
\end{figure}

We performed DMRG calculations of the triplet state of 1,3-dimethyl-imidazole-2-ylidene with an active space of 16 electrons and
28 orbitals. Hence, this is the first DMRG study to report CI coefficients for an active space
that cannot be treated by conventional CAS-type methods, which would span a complete many-particle
Hilbert space of dimension $8 \times 10^{12}$. The dimension of the DMRG system states was set to $m=60$, 80, 100, and 140
and one DMRG sweep comprises 25 microiteration steps. As orbital basis, natural orbitals from a CASSCF calculation employing
an active space of 12 electrons and 12 orbitals \cite{werner3,werner,werner2}
were taken. The CASSCF calculations as well as the calculation of the one-electron and two-electron integrals
were performed with the \textsc{Molpro} program package \cite{molpro} using Dunning's cc-pVTZ basis set
for all atoms \cite{dunning,dunning2}. We have monitored the CI coefficients during the variational
DMRG optimization for the $m=60$ DMRG calculation.

The DMRG ground state energies corresponding to different $m$ values and the CAS(12,12) reference
are given in Table \ref{tab:energies}. The DMRG ground state energy decreases systematically for
an increasing number of DMRG states $m$. From $m=80$ onwards, the DMRG energy is lower than the CAS(12,12)
reference energy as one would expect in view of the larger active space.
In this particular case, one may obtain lower electronic energies from a coupled-cluster or M{\o}ller--Plesset calculation 
and therefore we should emphasize that the Arduengo carbene serves simply as an example to explore the CI coefficients
of an CASCI-type wave function.

\begin{table}[h]
\caption{Ground state energies for the Arduengo carbene in Hartree for CAS(12,12) and DMRG calculations
employing a different number $m$ of DMRG basis states.}\label{tab:energies}
{
\begin{center}
\begin{tabular}{lc}\hline \hline
Method & $E$/Hartree \\\hline
CAS(12,12) & $-$302.933 197 \\ \hline
DMRG($m=60$) & $-$302.930 459 \\
DMRG($m=80$) & $-$302.933 253 \\
DMRG($m=100$) & $-$302.935 120 \\
DMRG($m=140$) & $-$302.937 288
\\ \hline
\hline
\end{tabular}
\end{center}
}
\end{table}
For $m=60$, the DMRG total electronic energy decreases considerably during the first four sweeps of the algorithm,
i.e., during the first 100 microiteration steps (see Fig.\ \ref{fig:energy}, in which the microiteration
steps from 51 to 250 are shown).
During these iterations the CI coefficients also change substantially.

\begin{figure}[h]
\centering
\includegraphics[width=0.9\linewidth]{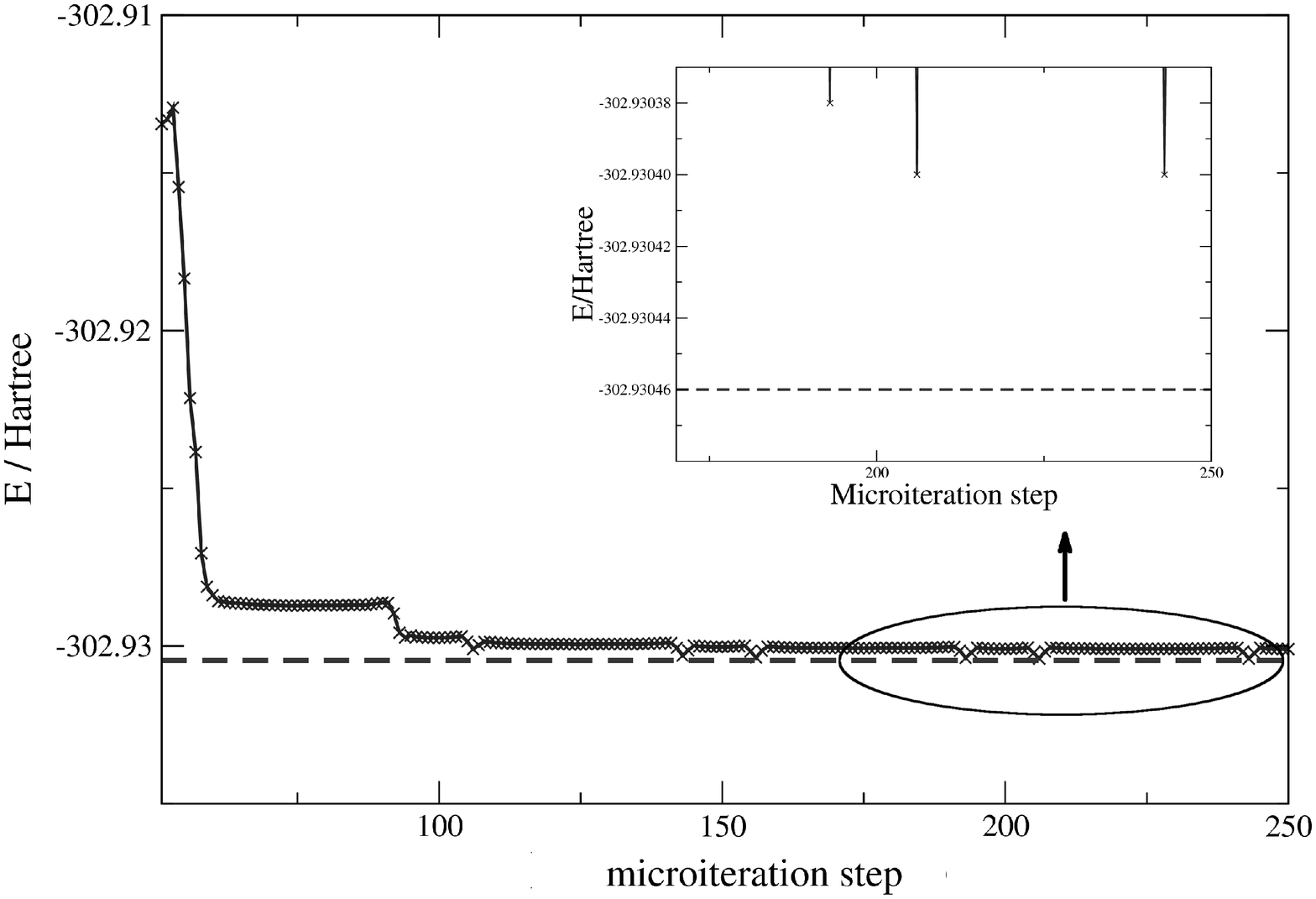}
\caption{Convergence of the Arduengo-carbene DMRG energy for $m=60$ displayed from the 51$^{\rm st}$ to the 250$^{\rm th}$
DMRG($m=60$) microiteration steps. The dashed red line corresponds to the energy of the final converged total state
which is reached after 50 sweeps, i.e.\ 1'250 microiterations.
}\label{fig:energy}
\end{figure}

The convergence behavior of the DMRG energy relates to the convergence of the CI coefficients. To
demonstrate this, we depict eight configurations corresponding to CI coefficients with largest weights present
in the final ground state wave function for the DMRG ($m=60$) calculation, denoted by SD1 to SD8,
where SD1 represents the Hartree--Fock reference
configuration. The evolution of the absolute value of the CI coefficients is shown in Fig.\ \ref{fig:sd}.
The total ground state is dominated by the reference configuration SD1 and contains several configurations
with small CI coefficients ($\leq 0.13$). The fast drop in energy during the first ten microiteration
steps of the third sweep involves the largest change of CI weights, where more configurations with
small CI coefficients are picked up. Afterwards, the DMRG energy reaches
a plateau where no changes in CI weights can be observed simultaneously, followed by a second, but smaller
decrease in energy at the same partitioning of the orbitals into system and environment as during the first energy jump.
Here, another significant change in CI weights can be observed. Then, both the DMRG energy and
the CI weights gradually converge to their corresponding final values.
\begin{figure}[h]
\centering
\includegraphics[width=0.6\linewidth]{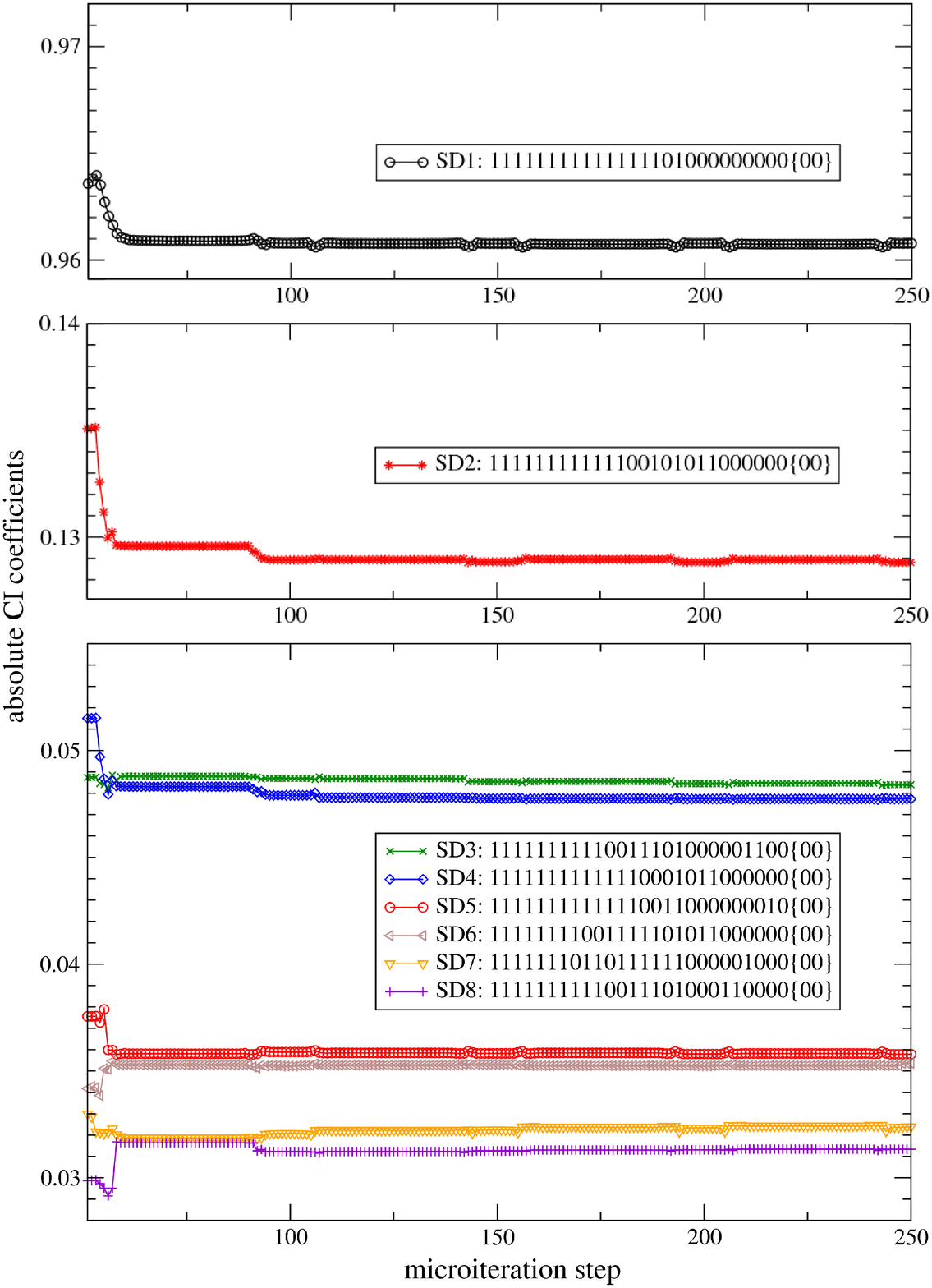}
\caption{Evolution of the absolute value of the Arduengo-carbene CI coefficients of selected configurations during the
DMRG microiteration steps. The binary numbers in the legend correspond to the occupation number vectors
in the spin orbital basis. Two consecutive spin orbitals represent one site 
of the lattice, where the first one corresponds to a
spin-up state and the second one to a spin-down state, respectively. Only the first 13 lattice sites
are shown, for all other lattice sites the spin orbitals are unoccupied which we abbreviated by \{00\}.
}\label{fig:sd}
\end{figure}
In Table \ref{tab:cis}, the CI coefficients corresponding to the eight monitored configurations of the total
ground state wave function are shown for different $m$ values. In general, the differences in CI weights for
an increasing number of DMRG states $m$ are minor and are up to approximately three orders of magnitude smaller
than the reference CI coefficients. 
It is clear that the electronic structure in such a single-reference case is dominated by 
the Hartree--Fock determinant and we shall focus on the contribution of the determinants with smaller CI weight.
Changes can be observed for SD4 (10\%), SD5 and SD6 (both 5\%).
In particular, the increase in CI weights for larger $m$ exchanges the weighted order of SD3 and
SD4, while for all other Slater determinants the ordering stays unchanged. 
With increasing $m$, the CI weight of the Hartree--Fock reference configuration decreases slightly resulting in larger contributions
from different excited configurations.
We thus observe that small changes in the CI coefficients induce large changes in energy (7 mHartree). 

However, the general pattern of magnitude of the CI
coefficients remains similar. Hence, CI coefficients have the same order of magnitude
for calculations with different $m$ values. 
As the CI coefficients do not change considerably when $m$ is increased, it is sufficient to start the sampling for
small $m$-DMRG calculations, and then refine the sampled Hilbert space for larger $m$. 
This makes the sampling procedure more efficient.
However, it is obvious that $m$ must not be too small, especially in multi-reference cases.

\begin{table}[h]
\caption{CI coefficients of the total ground state wave function of the Arduengo carbene
calculated for different $m$ values. The binary numbers correspond to the occupation number vectors
in the spin orbital basis. Two spin orbitals represent one site of the lattice. Only the first 13 lattice sites
are shown, for all other lattice sites the spin orbitals are unoccupied which we abbreviated by \{00\}.}\label{tab:cis}
{
\begin{center}
\begin{tabular}{lcccc}\hline \hline
\# & SD & $m=60$ & $m=80$ & $m=100$ \\\hline
1&11111111111111101000000000\{00\} & 0.960 540 & 0.958 072 & 0.957 032 \\
2&11111111111100101011000000\{00\} & 0.129 123 & 0.137 996 & 0.138 337 \\
3&11111111110011101000001100\{00\} & 0.048 283 & 0.048 724 & 0.048 635 \\
4&11111111111110001011000000\{00\} & 0.047 771 & 0.052 114 & 0.052 272 \\
5&11111111111110011000000010\{00\} & 0.035 522 & 0.035 752 & 0.037 352 \\
6&11111111001111101011000000\{00\} & 0.035 301 & 0.036 879 & 0.037 100 \\
7&11111110110111111000001000\{00\} & 0.032 598 & 0.032 291 & 0.032 384 \\
8&11111111110011101000110000\{00\} & 0.031 347 & 0.031 788 & 0.031 964
\\ \hline
\hline
\end{tabular}
\end{center}
}
\end{table}

We also studied the efficiency of our SR-CAS algorithm for different convergence criteria with respect to
the total DMRG ground state. For decreasing COM values the calculated energy 
for the sampled configurations converges towards the total ground state energy for a given number
of DMRG basis states (see Fig.\ \ref{fig:tre3}). Already for
${\rm COM}=0.000992643$, we capture 99.98\% of the ground state energy ($\Delta E=46\rm mH$), while the sampled
Hilbert space contains less than 0.94 ppb of the total number of Slater determinants (see also Fig.\ \ref{fig:tre1}).
A further decreasing of the COM to 9.98928$\times 10^{-5}$ captures 99.998\% of the total ground state energy
($\Delta E=4.6\rm mH$) in a Hilbert space spanned by only 70'916 Slater determinants (8.67 ppb
of the total Hilbert space). Our SR-CAS algorithm is thus capable of capturing the most important configurations
of the full Hilbert space while keeping the dimension of the CI-type expansion small.

It would also be interesting to study the number of determinants that needs to be considered for a given
accuracy in electronic energy with increasing size of the CAS. This, however, has already been studied by
Surj{\'a}n and co-workers who showed that the relative portion of important determinants shrinks 
with increasing size of the CAS \cite{roli08}, which is an advantage for our SR-CAS algorithm.

In general, we observe that the energy
corresponding to the sampled configurations is sensitive to the threshold value chosen. This energy
expectation value can serve as an additional convergence criterion. Hence, for an accurate
representation of the wave function, the threshold value can be reduced systematically until the desired accuracy
of the energy corresponding to the sampled configurations is reached. 

\begin{figure}[h]
\centering
\includegraphics[width=0.7\linewidth]{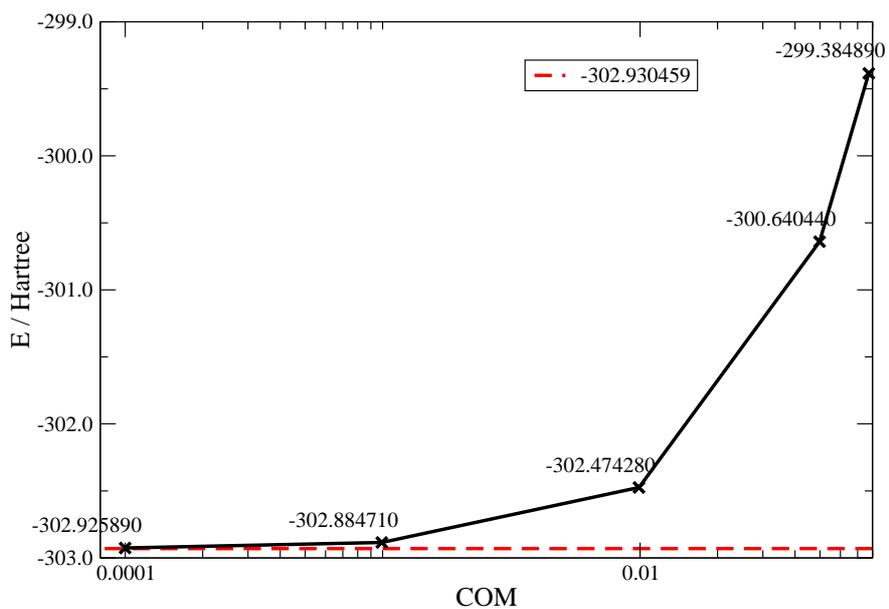}
\caption{Arduengo-carbene electronic energy corresponding to the sampled configurations for different COM values in Hartree.
The dashed red line represents the fully converged DMRG($m=60$) ground state energy.
}\label{fig:tre3}
\end{figure}

\begin{figure}[h]
\centering
\includegraphics[width=0.7\linewidth]{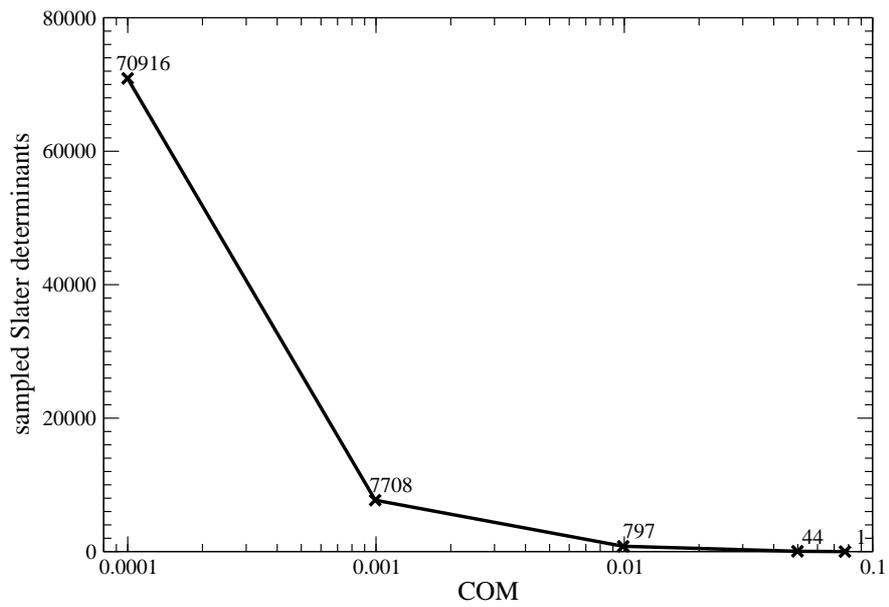}
\caption{Number of sampled Slater determinants for the DMRG($m=60$) calculation on the Arduengo carbene.
The total Hilbert space has a dimension of $8\times10^{12}$.
}\label{fig:tre1}
\end{figure}

\section{Conclusions\label{sec:conclusion}}
In this work, we presented a new method for the construction of CASCI-type wave functions
for molecular systems comprising a large active space.
The first study of a decomposition of DMRG basis states into Slater determinants \cite{gerrit_sd} suffered
from an exponential scaling and could therefore be applied to
a small active space only. Here, we significantly improved the decomposition scheme by
calculating the CI coefficients directly from matrix product states. Thereby, we benefit from the
polynomial scaling of the DMRG algorithm so that wave functions of large active spaces can
easily be analyzed in terms of the CI language. Since we are in general interested in the final
ground or excited state wave function, only the transformation matrices of the last two sweeps of the
algorithm need to be stored so that the additional computational cost and memory requirements are
negligible.
The fact that the complete determinant basis cannot be explicitly constructed for a large CAS necessitates
the introduction of the SR-CAS algorithm to pick up all determinants required to achieve a given
accuracy in the electronic energy. In turn, this sampling algorithm produces a minimal CI expansion
for a desired accuracy, which is surprisingly close to the full DMRG reference energy.

To enhance the efficiency of the sampling in the SR-CAS algorithm, an initial sampling of the full
Hilbert space can be performed for a small value $m$ of renormalized DMRG block states. These sampled
configurations can then
be used as input for large-$m$ calculations as one may expect the CI coefficients
to be of similar weight for different dimensions of DMRG system states. Provided that the initial DMRG calculation
did not employ a too small value for $m$, this should hold for single- and multi-reference cases.
In other words, our
procedure exploits the sparsity of large-scale CI problems in a most efficient manner. 
We have explicitly shown for the Arduengo carbene that only 70'916 out of $8\times10^{12}$ Slater
determinants are necessary for
a reliable CI expansion. Our analysis suggests that only a comparatively small number of
Slater determinants of the entire $N$-particle Hilbert space has to be considered to construct a very
efficient CASCI-type wave function.

Finally, we should emphasize that the presented SR-CAS algorithm can also be used to reconstruct
CI-type wave functions
for any tensor network state like complete-graph tensor networks \cite{cgtnarxiv2010} and correlator
product states \cite{cps}. Such a CI-type
expansion would facilitate the understanding and analysis of otherwise complicated wave functions.

\section*{Acknowledgments}

We gratefully acknowledge financial support by a TH-Grant (TH-26 07-3) from ETH Zurich.
KB acknowledges a Chemiefonds scholarship of the Fonds der Chemischen Industrie.


\begin{thebibliography}{10}

\bibitem{jorgensen}
T.~Helgaker, P.~J\o{}rgensen, and J.~Olsen,
\newblock {\em {Molecular Electronic-Structure Theory}},
\newblock Wiley, 2000.

\bibitem{Olsen1990}
J.~Olsen, P.~J\o{}rgensen, and J.~Simons,
\newblock Chem. Phys. Lett. {\bf 169}, 463 (1990).

\bibitem{knowles}
P.~J. Knowles and N.~C. Handy,
\newblock J. Chem. Phys. {\bf 91}, 2396 (1989).

\bibitem{knowles_fci}
P.~J. Knowles,
\newblock Chem. Phys. Lett. {\bf 155}, 513 (1989).

\bibitem{harrison}
R.~J. Harrison,
\newblock J. Chem. Phys. {\bf 94}, 5021 (1991).

\bibitem{luzanov}
A.~V. Luzanov, A.~L. Wulfov, and V.~O. Krouglov,
\newblock Chem. Phys. Lett. {\bf 197}, 614 (1992).

\bibitem{Mitrushenkov}
A.~O. Mitrushenkov,
\newblock Chem. Phys. Lett. {\bf 217}, 559 (1994).

\bibitem{feller}
D.~Feller,
\newblock J. Chem. Phys. {\bf 98}, 7059 (1993).

\bibitem{greer}
J.~C. Greer,
\newblock J. Chem. Phys. {\bf 103}, 1821 (1995).

\bibitem{greer98}
J.~C. Greer,
\newblock J. Comput. Phys. {\bf 146}, 181 (1998).

\bibitem{markus_chimia_2009}
M.~Reiher,
\newblock Chimia {\bf 63}, 140 (2009).

\bibitem{marti2010pccp}
K.~H. Marti and M.~Reiher,
\newblock Phys. Chem. Chem. Phys. {\bf {13}}, 6750 (2011),

\bibitem{markus_fd2}
M.~Podewitz, M.~T. Stiebritz, and M.~Reiher,
\newblock Faraday Discuss. {\bf 148}, 119 (2011).

\bibitem{white}
S.~R. White,
\newblock Phys. Rev. Lett. {\bf 69}, 2863 (1992).

\bibitem{PhysRevLett.68.3487}
S.~R. White and R.~M. Noack,
\newblock Phys. Rev. Lett. {\bf 68}, 3487 (1992).

\bibitem{dmrg_chan}
G.~K.-L. Chan, J.~J. Dorando, D.~Ghosh, J.~Hachmann, E.~Neuscamman, H.~Wang,
  and T.~Yanai,
\newblock {An Introduction to the Density Matrix Renormalization Group Ansatz
  in Quantum Chemistry},
\newblock in {\em {Frontiers in Quantum Systems in Chemistry and Physics}},
  edited by S.~Wilson, P.~J. Grout, J.~Maruani, G.~Delgado-Barrio, and
  P.~Piecuch, volume~{18} of {\em {Prog. Theor. Chem. Phys.}}, pages {49--65},
  Dordrecht, {2008}, Springer,
\newblock arXiv:0711.1398v1 [cond-mat.str-el].

\bibitem{marti2010b}
K.~H. Marti and M.~Reiher,
\newblock Z. Phys. Chem. {\bf 224}, 583 (2010).

\bibitem{chan_water}
G.~K.-L. Chan and M.~Head-Gordon,
\newblock J. Chem. Phys. {\bf 118}, 8551 (2003).

\bibitem{oestlund95}
S.~\"Ostlund and S.~Rommer,
\newblock Phys. Rev. Lett. {\bf 75}, 3537 (1995).

\bibitem{oestlund97}
S.~Rommer and S.~\"Ostlund,
\newblock Phys. Rev. B {\bf 55}, 2164 (1997).

\bibitem{gerrit_sd}
G.~Moritz and M.~Reiher,
\newblock J. Chem. Phys. {\bf 126}, 244109 (2007).

\bibitem{mitr03}
A.~O. Mitrushenkov, R.~Linguerri, P.~Palmieri, and G.~Fano,
\newblock J. Chem. Phys. {\bf 119}, 4148 (2003).

\bibitem{schollwoeck}
U.~Schollw\"ock,
\newblock Rev.\ Mod.\ Phys. {\bf 77}, 259 (2005).

\bibitem{marti2008}
K.~H. Marti, I.~{Malkin Ond{\`i}k}, G.~Moritz, and M.~Reiher,
\newblock J. Chem. Phys. {\bf 128}, 014104 (2008).

\bibitem{cgtnarxiv2010}
K.~H. Marti, B.~Bauer, M.~Reiher, M.~Troyer, and F.~Verstraete,
\newblock New J. Phys. {\bf 12}, 103008 (2010),
\newblock arXiv:1004.5303v1 [physics.chem-ph].

\bibitem{sandvik2007}
A.~W. Sandvik and G.~Vidal,
\newblock {Phys. Rev. Lett.} {\bf {99}}, {220602} ({2007}).

\bibitem{ozonstructure}
R.~Schinke and P.~Fleurat-Lessard,
\newblock J. Chem. Phys. {\bf 121}, 5789 (2004).

\bibitem{molpro}
{H.-J. {Werner}, P. J. {Knowles}, R. Lindh, F. R. Manby, M. Sch\"utz, P.
  Celani, T. Korona, A. Mitrushenkov, G. Rauhut, T. B. Adler, R. D. Amos, A.
  Bernhardsson, A. Berning, D. L. Cooper, M. J. O. Deegan, A. J. Dobbyn, F.
  Eckert, E. Goll, C. Hampel, G. Hetzer, T. Hrenar, G. Knizia, C. K\"oppl, Y.
  Liu, A. W. Lloyd, R. A. Mata, A. J. May, S. J. McNicholas, W. Meyer, M. E.
  Mura, A. Nicklass, P. Palmieri, K. Pfl\"uger, R. Pitzer, M. Reiher, U.
  Schumann, H. Stoll, A. J. Stone, R. Tarroni, T. Thorsteinsson, M. Wang and A.
  Wolf},
\newblock Molpro, version 2009.1, a package of \emph{ab initio} programs, 2009,
\newblock see http://www.molpro.net.

\bibitem{dunning}
J.~T.~H.~Dunning,
\newblock J. Chem. Phys. {\bf 90}, 1007 (1989).

\bibitem{dunning2}
N.~B. Balabanov and K.~A. Peterson,
\newblock J. Chem. Phys. {\bf 123}, 064107 (2005).

\bibitem{dmrg_new}
G.~Moritz, K.~H. Marti, K.~Boguslawski, and M.~Reiher,
\newblock {\em {\textsc{Qc-Dmrg-ETH}, A Program for Quantum Chemical DMRG
  Calculations. { \rm Copyright 2007--2010, ETH Z\"urich}}}.

\bibitem{dreu2005}
A.~Dreuw and M.~Head-Gordon,
\newblock Chem. Rev. {\bf 105}, 4009 (2005).

\bibitem{dreu2006}
A.~Dreuw,
\newblock ChemPhysChem {\bf 7}, 2259 (2006).

\bibitem{neug2010}
J.~Neugebauer,
\newblock Phys. Rep. {\bf 489}, 1 (2010).

\bibitem{neug2011}
J.~Neugebauer,
\newblock {Orbital-Free Embedding Calculations of Electronic Spectra},
\newblock in {\em Recent Advances in Orbital-Free Density Functional Theory},
  page in press, World Scientific, Singapore, 2011.

\bibitem{legeza_molphys}
\"O.~Legeza, J.~R\"oder, and B.~A. Hess,
\newblock Mol. Phys. {\bf 101}, 2019 (2003).

\bibitem{arduengo_methyl}
A.~J. Arduengo, H.~V.~R. Dias, R.~L. Harlow, and M.~Kline,
\newblock J. Am. Chem. Soc. {\bf 114}, 5530 (1992).

\bibitem{werner3}
H.-J. Werner and W.~Meyer,
\newblock J. Chem. Phys. {\bf 74}, 5794 (1981).

\bibitem{werner}
H.-J. Werner and P.~J. Knowles,
\newblock J. Chem. Phys. {\bf 82}, 5053 (1985).

\bibitem{werner2}
P.~J. Knowles and H.-J. Werner,
\newblock Chem. Phys. Lett. {\bf 115}, 259 (1985).

\bibitem{roli08}
Z.~Rolik, {\'A}.~Szabados, and P.~R. Surj{\'a}n,
\newblock J. Chem. Phys. {\bf 128}, 144101 (2008).

\bibitem{marti2010a}
K.~H. Marti and M.~Reiher,
\newblock Mol. Phys. {\bf 108}, 501 (2010).

\bibitem{cps}
H.~Changlani, J.~Kinder, C.~Umrigar, and G.~Chan,
\newblock \rm arXiv:0907.4646v1 .

\bibitem{p86}
J.~P. Perdew,
\newblock Phys. Rev. B {\bf 33}, 8822 (1986).

\bibitem{b88}
A.~D. Becke,
\newblock Phys. Rev. A {\bf 38}, 3098 (1988).

\bibitem{adf}
G.~T. Velde, F.~M. Bickelhaupt, E.~J. Baerends, C.~F. Guerra, S.~J.~A.
  Van~Gisbergen, J.~G. Snijders, and T.~Ziegler,
\newblock J. Comput. Chem. {\bf 22}, 931 (2001),
\newblock see http://www.scm.com.

\bibitem{koni_diss}
K.~H. Marti,
\newblock {\em {New Electron Correlation Theories and Haptic Exploration of
  Molecular Systems}},
\newblock PhD thesis, {ETH Z{\"u}rich}, 2010.

\bibitem{lege03a}
{\"O}.~Legeza, J.~R\"oder, and B.~A. Hess,
\newblock Phys. Rev. B {\bf 67}, 125114 (2003).

\bibitem{lege04}
{\"O}.~Legeza and J.~S{\'o}lyom, 
\newblock Phys. Rev. B {\bf 70}, 205118 (2004).

\bibitem{peres1984}
A.~Peres,
\newblock Phys. Rev. A {\bf 30}, 1610 (1984).

\bibitem{zhou2008}
H.-Q. Zhou, R.~Orus, and G.~Vidal,
\newblock {Phys. Rev. Lett.} {\bf {100}}, {080601} ({2008}).

\bibitem{lege03c}
{\"O}.~Legeza and J.~S{\'o}lyom,
\newblock Phys. Rev. B {\bf 68}, 195116 (2003).

\bibitem{hanrath08}
M.~Hanrath,
\newblock Chem. Phys. Lett. {\bf 466}, 240 (2008).

\bibitem{hanrath09}
A.~Engels-Putzka and M.~Hanrath,
\newblock J. Mol. Struct.: THEOCHEM {\bf 902}, 59 (2009).

\bibitem{peyerimhoff74}
R.~J. Buenker and S.~D. Peyerimhoff,
\newblock {Theoret. Chim. Acta} {\bf {35}}, {33} ({1974}).

\bibitem{peyerimhoff75}
R.~J. Buenker and S.~D. Peyerimhoff,
\newblock {Theoret. Chim. Acta} {\bf {39}}, {217} ({1975}).

\bibitem{peyerimhoff77}
R.~J. Buenker, S.~D. Peyerimhoff, and W.~Butscher,
\newblock {Mol. Phys.} {\bf {35}}, {771} ({1977}).

\end{thebibliography}
\end{document}